\documentclass[%
aip,
apl,
 amsmath,amssymb,
 reprint,%
]{revtex4-1}

\usepackage{graphicx}
\usepackage{dcolumn}
\usepackage{bm}

\usepackage[utf8]{inputenc}
\usepackage[T1]{fontenc}
\usepackage{mathptmx}
\usepackage{etoolbox}
\usepackage{xcolor}
\usepackage{booktabs}
\usepackage{soul}
\usepackage{float}
\makeatletter
\def\@email#1#2{%
 \endgroup
 \patchcmd{\titleblock@produce}
  {\frontmatter@RRAPformat}
  {\frontmatter@RRAPformat{\produce@RRAP{*#1\href{mailto:#2}{#2}}}\frontmatter@RRAPformat}
  {}{}
}%
\makeatother
\let\oldtheta\theta
\renewcommand{\theta}{\ensuremath{\oldtheta}}
\begin{document}

\preprint{AIP/123-QED}

\title[Enhanced Performance of FeFET Gate Stack  via Heterogeneously co-doped Ferroelectric HfO$_2$ Films]{Enhanced Performance of FeFET Gate Stack  via Heterogeneously co-doped Ferroelectric HfO$_2$ Films}

\author{Shouzhuo Yang}
\email{shouzhuo.yang@ipms.fraunhofer.de}
\affiliation{Center Nanoelectronic Technologies CNT, Fraunhofer Institute for Photonic Microsystems IPMS, 01109 Dresden, Germany}
\affiliation{Institute of Solid-State Electronics, Faculty of Electrical and Computer Engineering, Technische Universität Dresden, 01069 Dresden, Germany}

\author{David Lehninger}%
\affiliation{Center Nanoelectronic Technologies CNT, Fraunhofer Institute for Photonic Microsystems IPMS, 01109 Dresden, Germany}

\author{Peter Reinig}%
\affiliation{Center Nanoelectronic Technologies CNT, Fraunhofer Institute for Photonic Microsystems IPMS, 01109 Dresden, Germany}

\author{Fred Sch\"{o}ne}
\affiliation{Center Nanoelectronic Technologies CNT, Fraunhofer Institute for Photonic Microsystems IPMS, 01109 Dresden, Germany}

\author{Raik Hoffmann}
\affiliation{Center Nanoelectronic Technologies CNT, Fraunhofer Institute for Photonic Microsystems IPMS, 01109 Dresden, Germany}

\author{Konrad Seidel}
\affiliation{Center Nanoelectronic Technologies CNT, Fraunhofer Institute for Photonic Microsystems IPMS, 01109 Dresden, Germany}

\author{Maximilian~Lederer}
\affiliation{Center Nanoelectronic Technologies CNT, Fraunhofer Institute for Photonic Microsystems IPMS, 01109 Dresden, Germany}

\author{Gerald Gerlach}
\affiliation{Institute of Solid-State Electronics, Faculty of Electrical and Computer Engineering, Technische Universität Dresden, 01069 Dresden, Germany}

\date{\today}
\begin{abstract}
In this work, we explore the impact of spatially controlled Zr and Al heterogeneous co-doping in HfO$_2$ thin films tailored for metal-ferroelectric-insulator-semiconductor (MFIS) gate stacks of ferroelectric field effect transistors (FeFETs). By precisely modulating the vertical arrangement of Zr and Al dopants during atomic layer deposition, we introduce deliberate compositional gradients that affect crystallization dynamics during subsequent annealing. This strategy enables us to systematically tune the phase evolution and domain nucleation within the ferroelectric layer, directly influencing device reliability and performance. From a structural perspective, our findings demonstrate that the phase composition of annealed HfO$_2$ films in MFIS stacks is primarily determined by the spatial arrangement of dopants. From an electrical perspective, we observe significant enhancement of remanent polarization and endurance of the gate stacks through heterogeneous co-doping, depending on the spatial arrangement of dopants. \\
\end{abstract}
\maketitle
\begin{figure}[b]
\includegraphics[width=\linewidth]{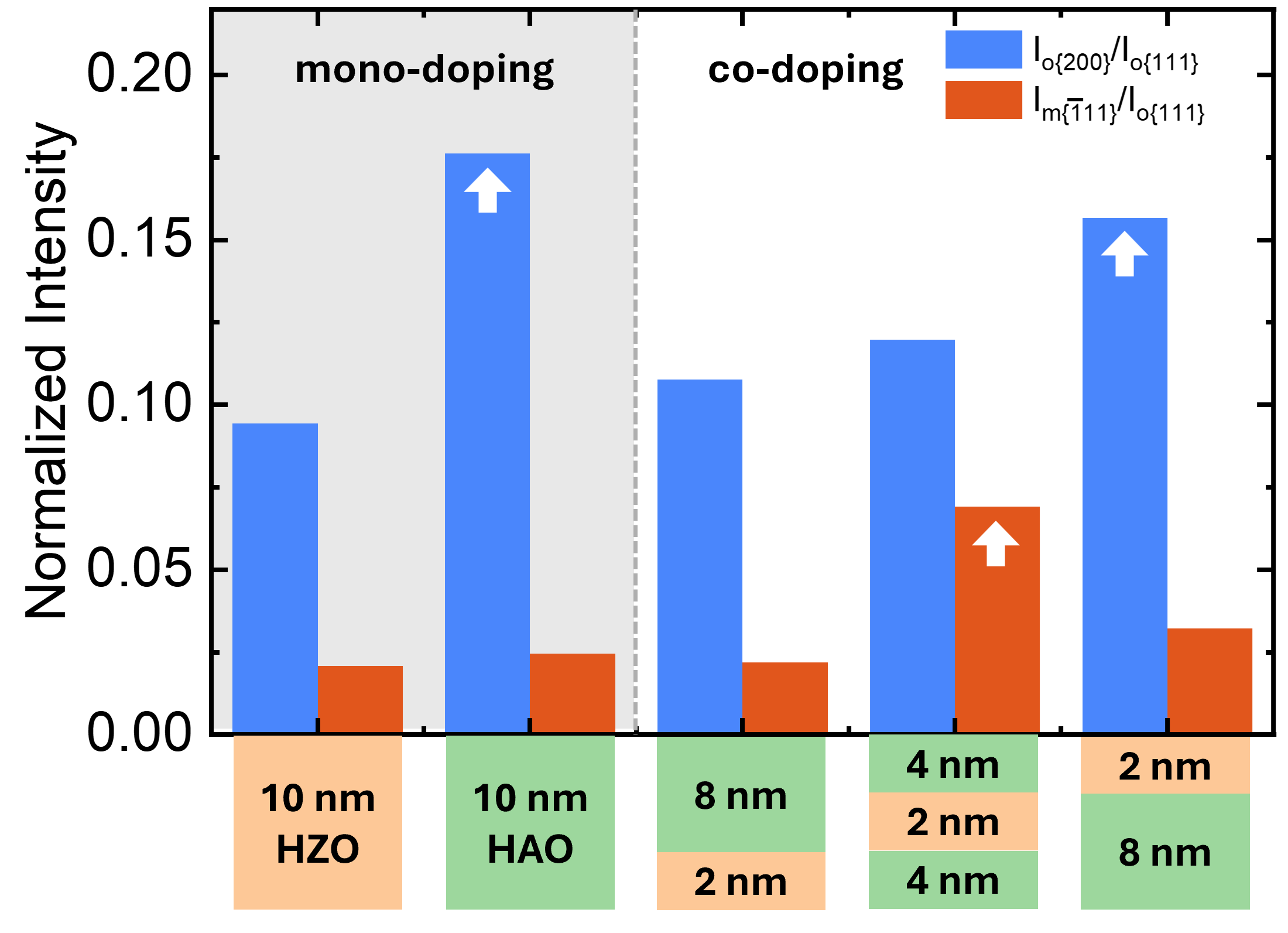}
\caption{Quantitative comparison of the GIXRD patterns among all samples with normalized intensity of the peaks of o(111), o\{200\} and m$(\bar{1}11)$.}
\label{XRD_Comparison}
\end{figure}

The von Neumann bottleneck~\cite{Park2023AM} of conventional memory devices have motivated the search for novel device architectures, such as ferroelectric field-effect transistors (FeFETs)~\cite{Hoffman2010AM}, that leverage the switchable polarization of doped HfO$_2$~\cite{Mueller2011APL, Mulaosmanovic2021NT, Schroeder2022NRM}. In particular, the metal-ferroelectric-insulator-semiconductor (MFIS) structure, which incorporates a ferroelectric layer into the gate stack, offers a promising pathway for integrating nonvolatile memory functions within mainstream CMOS logic platforms. The ability to stabilize robust ferroelectricity in ultra-thin HfO$_2$ films through appropriate dopant engineering is central to advancing FeFET technology for next-generation computing, i.e. for in-memory computing~\cite{Khan2020NE,Soliman2023NC} and neuromorphic computing~\cite{Thomann2023IRPS,Xiang2023AFM,Mehonic2024APLMat}.\\
\begin{figure}[b]
  \includegraphics[width=\linewidth]{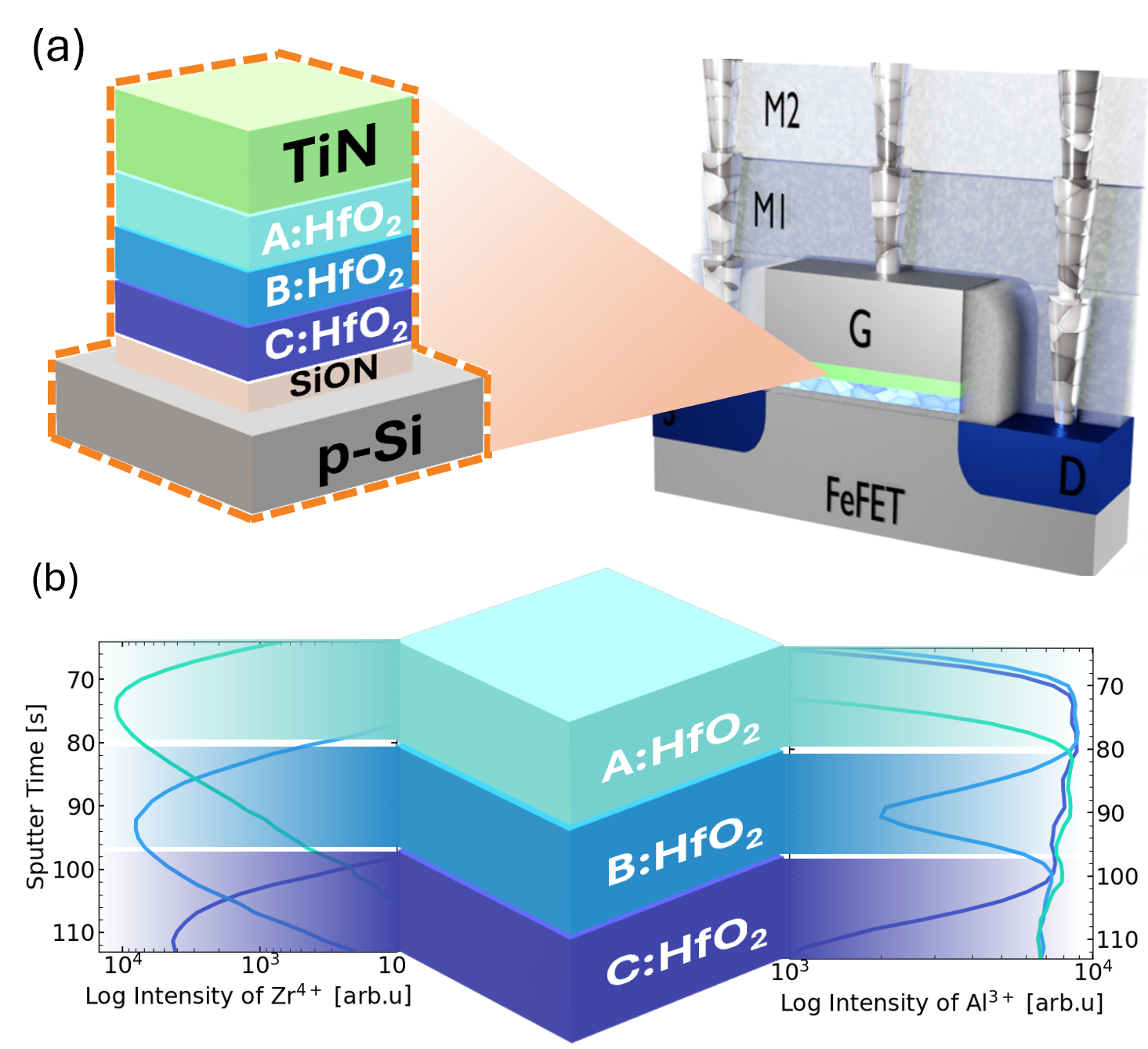} 
\caption{Schematic of a 1T-FeFET and its gate structure featuring a heterogeneously co-doped MFIS stack: separate layers (A/B/C) of HfO$_2$ individually doped with distinct elements through ALD and (b) ToF-SIMS profiles of Zr\textsuperscript{4+} (left) and Al\textsuperscript{3+} (right) ions of the three co-doped HfO$_2$ films.}
\label{Schematic_ToF-SIMS}
\end{figure}
HfO$_2$, a high-$\kappa$ dielectric material, has been established as a key component in advanced CMOS technology notes.~\cite{Houssa2006MSE} Since 2007, the ferroelectric HfO$_2$ has been discovered by introduction of dopants into HfO$_2$ thin films.~\cite{Heitmann2007USP, Boescke2011APL} It retains ferroelectricity in ultra-thin films down to 1~nm~\cite{Cheema2022AEM}, which makes it an ideal candidate with high scalability for CMOS-compatible FeFETs devices~\cite{Mueller2012VLSI, Trentzsch2016IEDM, Duenkel2017IEDM}. The doped HfO$_2$ thin films become polycrystalline after annealing, exhibiting orthorhombic (o), monoclinic (m), tetragonal (t), and cubic phases.~\cite{Boescke2011APLPhase,Park2019NanoS,Wei2021AFM,ZHOU2019CMS,Mueller2012NL} The prevailing opinion suggests that the doping induced ferroelectricity of HfO$_2$ thin films arises from the meta-stable non-centrosymmetric o-phase (Pca2$_1$)~\cite{Park2015AM}, whereas the m-phase is thermodynamically stable at room temperature~\cite{Boescke2011APL}. Therefore, the structural properties of ferroelectric HfO$_2$ thin films likely determine their electrical performance.~\cite{Sang2015APL,Lee2022JAP,Lederer2021ACSAEM}\\
Currently, the performance of HfO$_2$-based FeFETs is often hindered by reliability issues~\cite{Mulaosmanovic2021IPRS,Lederer2021MRSAdv,Ma2019IMW} such as small memory window, short endurance, and short retention time. These challenges necessitate the exploration of novel doping strategies to enhance the ferroelectric properties and overall reliability of HfO$_2$ films.\\
Building on our prior findings with heterogeneously co-doped metal-ferroelectric-metal (MFM) stacks~\cite{Yang2024APL}, we extend the heterogeneous co-doping concept to Hf$_x$Zr$_{1-x}$O$_2$/Hf$_x$Al$_{1-x}$O$_y$ (HZO/HAO) systems in MFIS stacks. By engineering the sequential placement of Zr- and Al-doped layers within the HfO$_2$ film, we intentionally modulate local crystallization behavior. This enables tailored phase stabilization and defect control during thermal processing, offering a means to optimize both ferroelectric properties and endurance specifically for MFIS device requirements.\\
This work aims to investigate the structural and electrical characteristics of Zr/Al co-doped HfO$_2$ films within the MFIS stack. By employing Atomic Layer Deposition (ALD) techniques, precise control over the doping concentration and location can be achieved in HfO$_2$ films, leading to optimized ferroelectric properties of FeFET gate stacks. Through comprehensive analysis utilizing Time-of-Flight Secondary Ion Mass Spectrometry (ToF-SIMS) and grazing incidence X-ray Diffraction (GIXRD), we will elucidate the impact of co-doping on the film's crystalline structure and ferroelectric performance. Furthermore, electrical characterization through fatigue measurements (FM) and Positive-Up Negative-Down (PUND) tests will provide insights into the endurance and leakage current behavior of the fabricated MFIS capacitors.\\
\begin{figure}[tb]
\includegraphics[width=\linewidth]{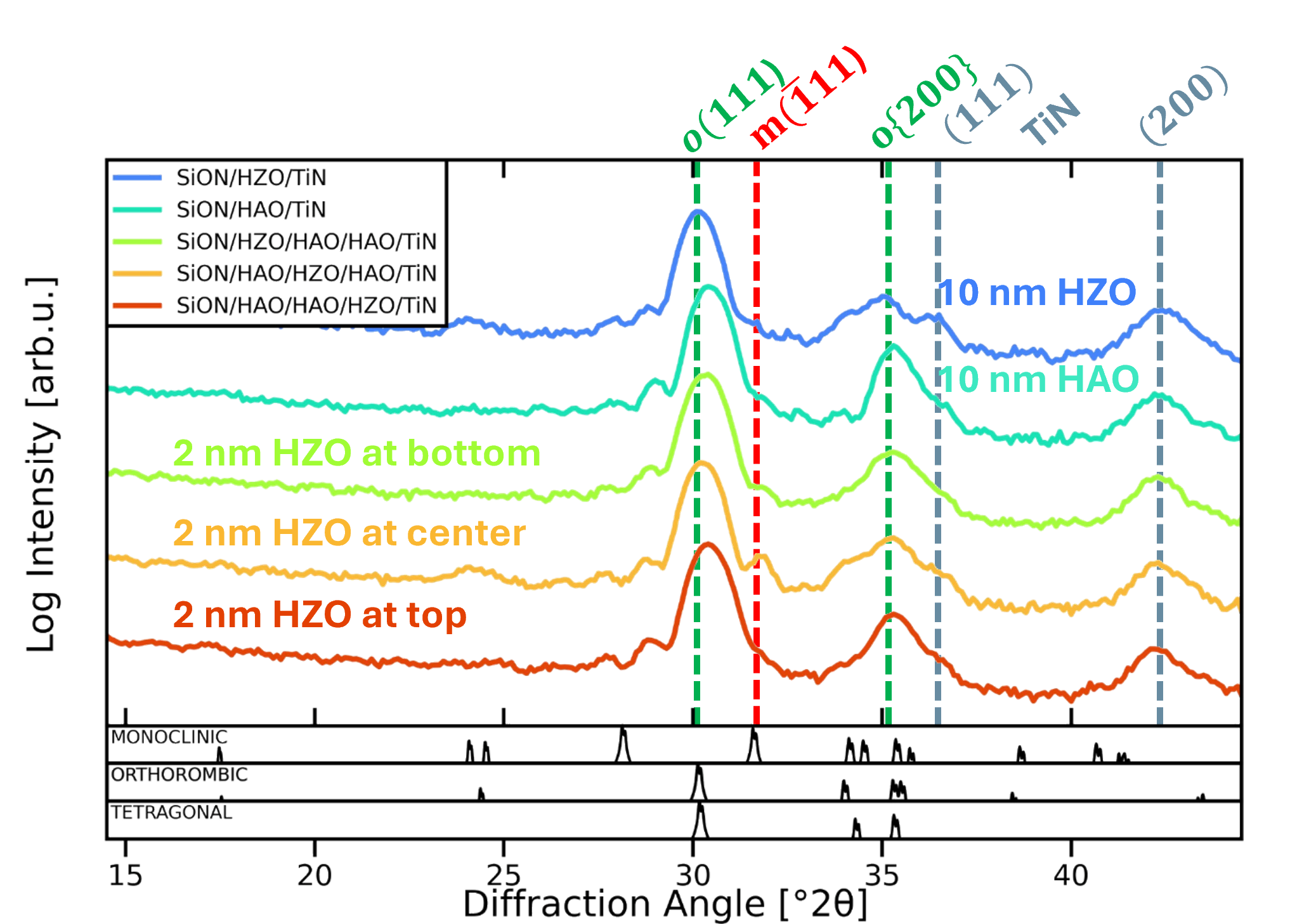}
\caption{GIXRD patterns of annealed samples of the experiment and control group with remarks of the peaks of o-phase, m-phase and TiN electrodes in green, red and gray, respectively.}
\label{GIXRD} 
\end{figure}
The investigated samples were fabricated in MFIS stacks with mono-doped and heterogeneously co-doped HfO$_2$ films with various dopants arrangement. Following the removal of the native oxide, the surface of highly p-doped Si wafers were chemically oxidized to form a SiO$_2$ insulator approx. 1~nm thick. Subsequently, it underwent a two-step rapid thermal process (RTP) in an ammonia atmosphere, followed by exposure to oxygen-nitrogen gas, resulting in a SiO$_2$/SiON interfacial layer (IL) approximately 2 nm thick. 10~nm co-doped HfO$_2$ thin films were deposited via ALD at 250~°C using HfCl$_4$ as the Hf precursor, metal-organic precursors for Zr and Al, and H$_2$O as the oxidizer. Zr and Al atoms were introduced into distinct layers of the HfO$_2$ film by tuning the sequence of precursor cycles, as shown in FIG.~\ref{Schematic_ToF-SIMS}a.\\
For the experiment group, MFIS stacks with co-doped HfO$_2$ films were doped as HZO and HAO in configurations of IL/HZO/HAO/HAO, IL/HAO/HZO/HAO, and IL/HAO/HAO/HZO, corresponding to Zr doping at the bottom (C), center (B), and top (A) layers, respectively. The control group included MFIS stacks with mono-doped HZO and HAO films, each containing only a single dopant species. All samples maintained optimal doping ratios of Zr:Hf = 1:1 for HZO and Al:Hf = 1:30 for HAO. A 10~nm TiN capping layer was deposited using physical vapor deposition (PVD), followed by annealing at 800~°C for 20~s using RTP. The configuration of all samples can be found in FIG.~\ref{XRD_Comparison}.\\
The spatial distribution of Zr and Al dopants throughout the HfO$_2$ films was characterized using time-of-flight secondary ion mass spectrometry (ToF-SIMS), confirming the successful realization of distinct compositional gradients as programmed by the ALD cycle sequence. As shown in FIG.~\ref{Schematic_ToF-SIMS}b, the depth profiles reveal effective dopant separation and diffusion behavior pertinent to the various stacking configurations under study.\\
Phase identification and texture analysis of the post-annealed HfO$_2$ films were performed using GIXRD within a 2$\theta$ window of 15° to 45°. This allowed us to correlate the emergence of orthorhombic and monoclinic phases with the vertical dopant arrangement, providing insights into how Zr/Al placement influences phase competition and preferred crystallographic orientations in MFIS devices.\\
On the one hand, FIG.~\ref{GIXRD} indicates that the GIXRD patterns of all five samples in both the experiment and control groups display peaks corresponding to both the o- and m-phases. However, most of the grains in all polycrystalline mono- and co-doped films are crystallized in the o-phase, with a dominant orientation along the (111) plane, typically referred to as o(111).\\
On the other hand, when HZO and HAO layers are deposited to create heterogeneously co-doped HfO$_2$ films, the GIXRD patterns show a significant change in the intensity of the m$(\bar{1}11)$ peak. Our results show that situating the HZO near the center of the HfO$_2$ film intensifies monoclinic phase formation, as evidenced by the pronounced m$(\bar{1}11)$ diffraction peak in FIG.~\ref{GIXRD}. This effect arises from the interplay between local crystallization temperatures and interfacial constraints~\cite{Yang2024APL}, where HZO facilitates nucleation at significantly lower temperatures down to 300~°C~\cite{Lehninger2020ISAF,Lehninger2021AEM}, compared to HAO at temperatures above 650~°C~\cite{Mueller2012AFM}. The HfO$_2$ film started crystallizing from HZO at the center without the confinement of top/bottom interfaces, leading to an increased monoclinic fraction~\cite{Yang2024APL}. In contrast, confining HZO layers to the top or bottom interface enhances orthorhombic phase stabilization due to stronger capping effects from adjacent materials~\cite{Chernikova2015MircoEng}.Thus, it is demonstrated at the phenomenological level that controlling dopants via heterogeneous co-doping significantly affects the crystallographic phase composition in HfO$_2$ films incorporated into MFIS stacks.\\
It is noteworthy that the pronounced shift of the o(111) peak in films containing HAO is likely attributed to tensile lattice strain~\cite{Dolabella2022SM} and/or the existence of higher symmetry phases~\cite{Mueller2012NL}.\\
\begin{figure}[tb]
\includegraphics[width=\linewidth]{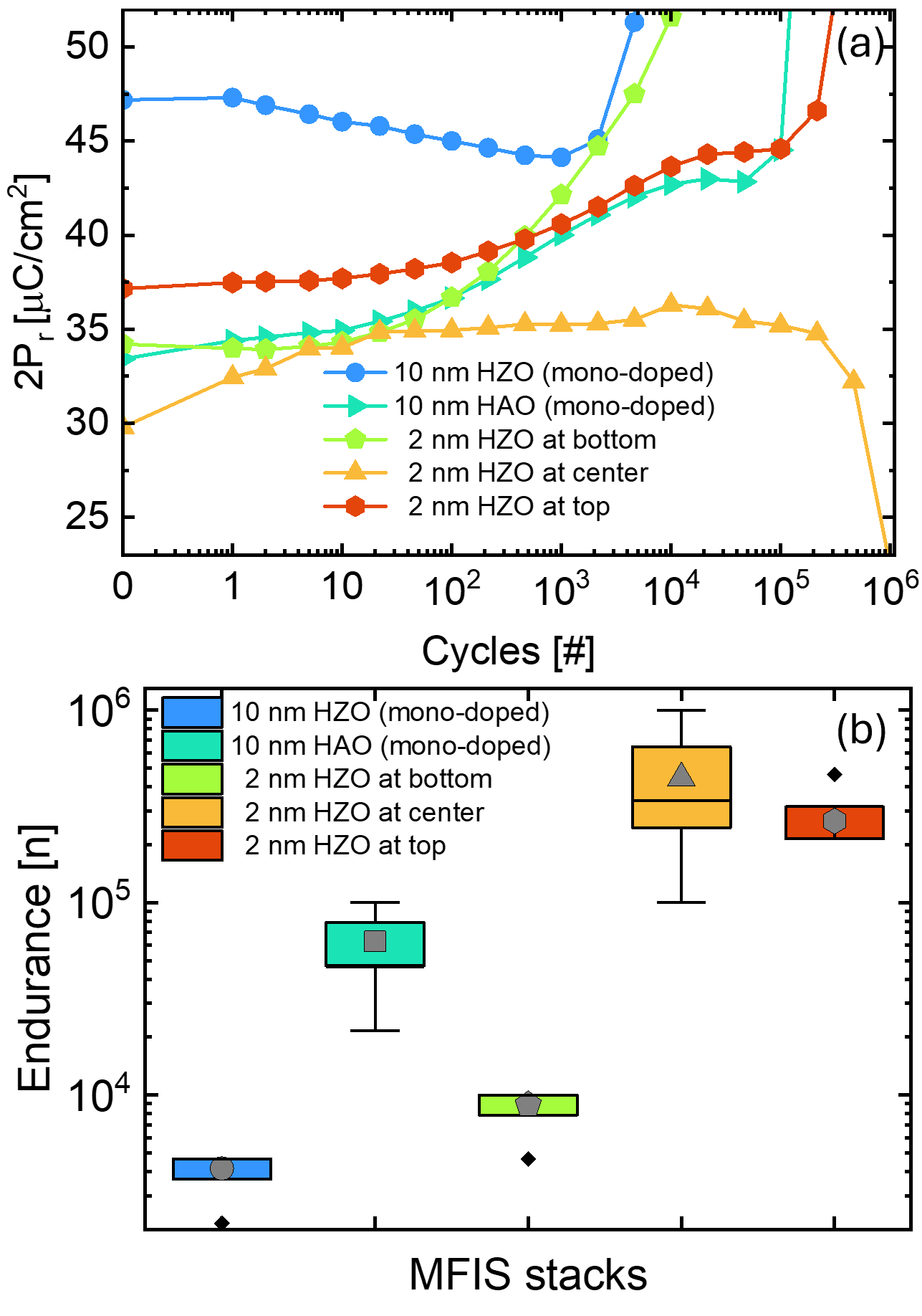}
\caption{(a) Fatigue measurements on MFIS capacitors in dependence on cycle count, (b) endurance extracted from statistical measurements.}
\label{FM} 
\end{figure}
\begin{figure}[hb]
\includegraphics[width=\linewidth]{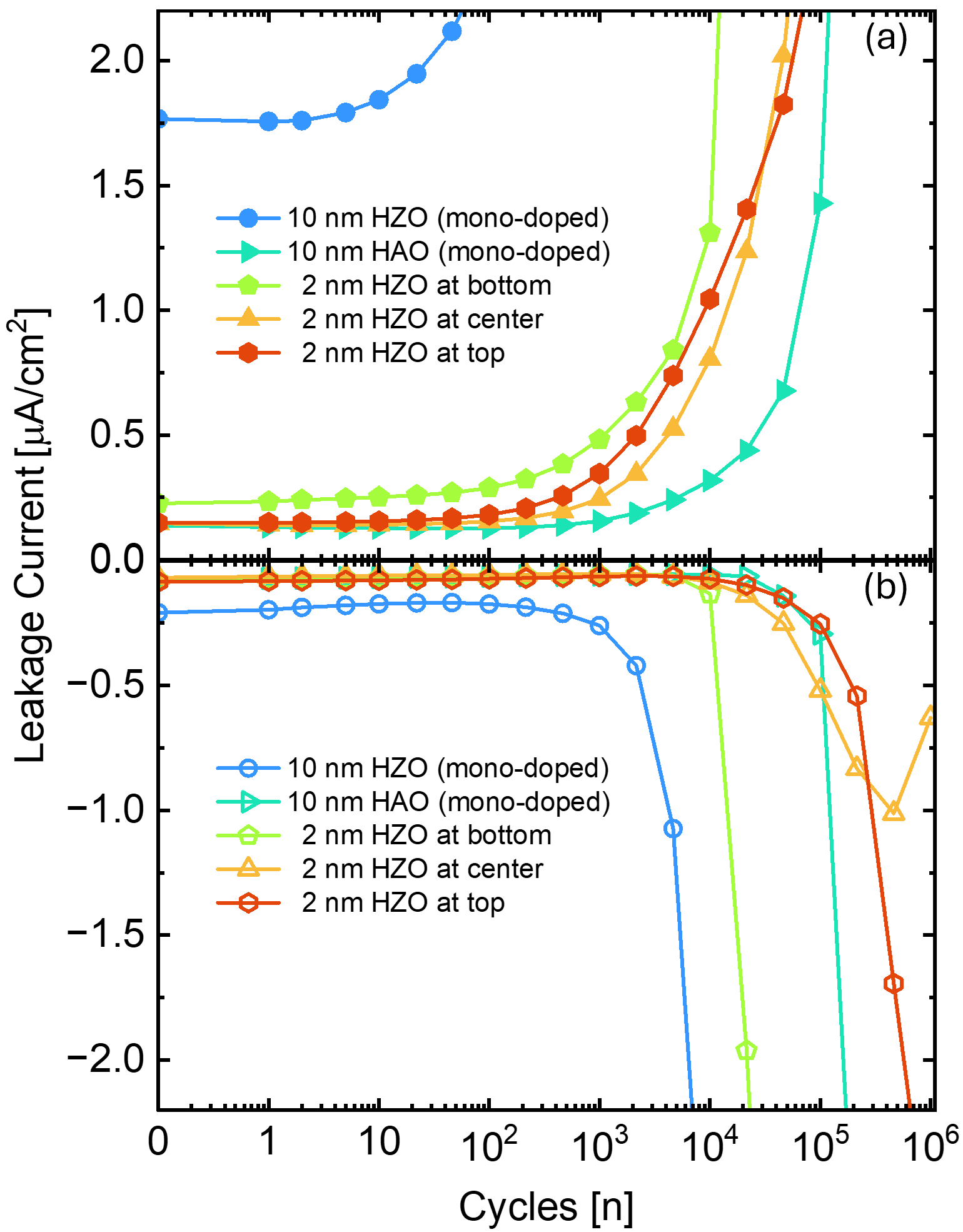}
\caption{(a) Extracted leakage current from PUND measurements while cycling (a) in positive direction and (b) in negative direction.}
\label{Leakage} 
\end{figure}
\begin{figure*}[htbp]
\includegraphics[width=\linewidth]{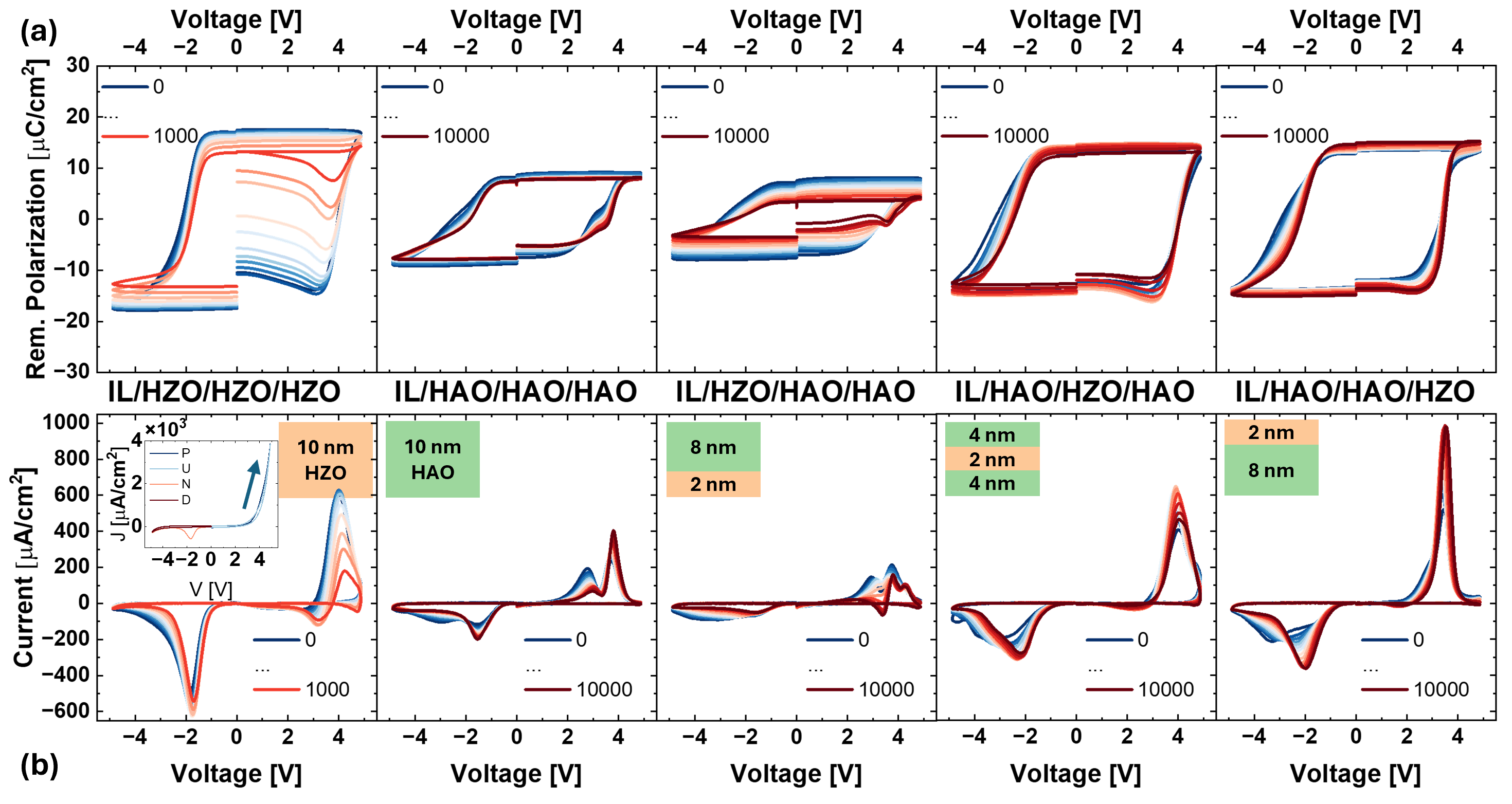} 
\caption{(a) P-V and (b) J-V loops extracted from PUND measurements while cycling of MFIS capacitors with Zr/Al mono- and co-doped HfO$_2$ films. Inset: J-V curves corresponded to each pulse of PUND measurements after 10$^3$ cycles.}
\label{PE_loops}
\end{figure*}
\begin{figure}[b]
\includegraphics[width=\linewidth]{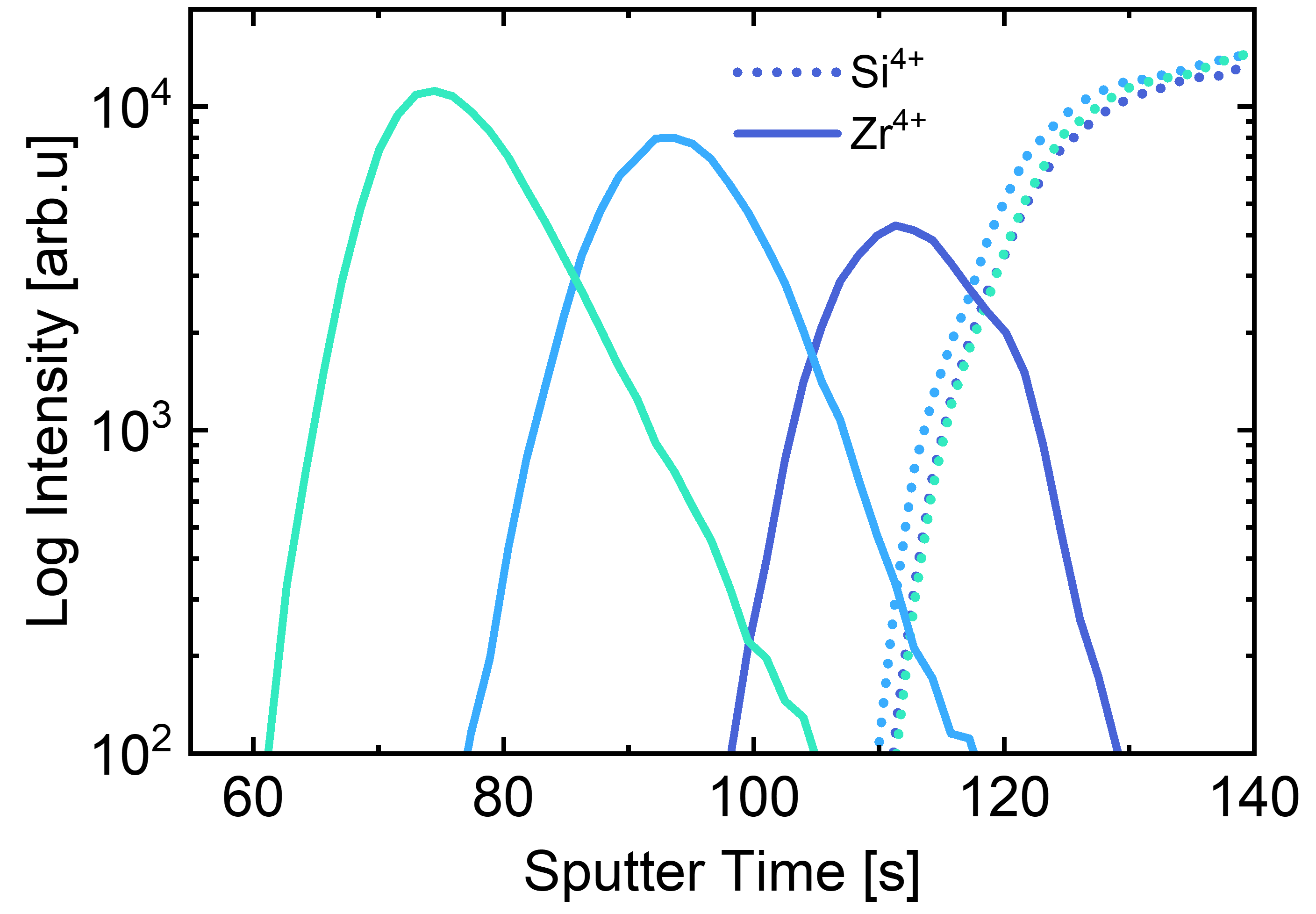}
\caption{Vertical distribution of Zr\textsuperscript{4+} ions measured by ToF-SIMS in the stacks with co-doped HfO$_2$ films. The overlap of the purple solid and dashed lines is highlighted, corresponding to the stack with SiON/HZO interface.}
\label{Diffusion} 
\end{figure}
FIG.~\ref{XRD_Comparison} provides a quantitative comparison of the main o-/m-peaks of GIXRD patterns shown in FIG.~\ref{GIXRD}. GIXRD patterns were quantitatively analyzed by WiRE\textsuperscript{TM} 5.1 software (Renishaw plc). The data processing procedure consisted of the following steps: first, truncation of the XRD scan angle range to 20°- 40°; second, baseline subtraction using a polynomial baseline correction; third, normalization of the XRD datasets to the maximum peak intensity. Finally, individual GIXRD peaks were analyzed using a curve-fitting algorithm with a combined Gaussian–Lorentzian line shape. This procedure allowed for the extraction of the peak center positions, full widths at half maximum (FWHM), and peak intensities. No smoothing was applied to the measurement data.Subsequently, the intensities of the o\{200\} and m$(\bar{1}11)$ peaks (I$_{o\{200\}}$ and I$_{m(\bar{1}11)}$) are normalized to the intensity of the o(111) peak (I$_{o(111)}$). From the intensity of the o\{200\} lines, it can be observed that it is strongly influenced by the position of the HZO layer. Especially the co-doped sample with HZO at the top shows an increased o\{200\} intensity, which is often associated with an increased crystallographic texture or semi-epitaxial growth, also referred to as templating effect~\cite{Lederer2021APL,Chae2022ACSAM}. Surprisingly, the mono-doped HAO sample shows a similar effect. In contrast to MFM stacks~\cite{Yang2024APL}, MFIS stacks do not have a pre-textured bottom electrode.\\
In order to assess the endurance of co-doped MFIS stacks, FM with $\pm$5~V pulse cycling at 10~kHz was conducted on capacitors as depicted in Fig.~\ref{FM}. It can be noticed that the 10 nm HZO mono-doped film exhibits the highest remanent polarization (2P$_r$) but suffers from the shortest endurance. Conversely, the 10 nm HAO mono-doped film is observed to have a lower 2P$_r$ but longer endurance. Interestingly, comparable 2P$_r$ values are exhibited by the three co-doped stacks, yet quite different endurance performances are observed. For instance, the capacitor experiences rapid fatigue when HZO is placed at the bottom. In contrast, the highest endurance is observed when HZO is positioned at the center of the film. This suggests that, while the endurance of capacitors can be significantly enhanced through co-doping, the spatial distribution of dopants is crucial.\\
Referring back to the GIXRD results in Fig.~\ref{GIXRD}, positioning HZO at the center of the film induces a greater amount of the m-phase. This enhances the stack's resistance to cycling~\cite{Song2022AEM}, albeit at the expense of reducing 2P$_r$.\\
Additionally, as demonstrated by the statistical results in Fig.~\ref{FM}b, the endurance of capacitors significantly diminishes when HZO is positioned at the bottom, applicable to both mono-doped and co-doped HfO$_2$ films. In other words, the direct contact between IL and HZO (occurrence of SiON/HZO interface) substantially compromises the strength of the MFIS stack against cycling stress.\\
To investigate the underlying causes, PUND measurements were also carried out at $\pm$5~V. As illustrated in Fig.~\ref{Leakage}, the leakage of MFIS capacitors was tracked during cycling. It is evident that the mono-doped HZO stack consistently demonstrates a much higher leakage current than other stacks, particularly in the positive direction. This can be attributed to the high annealing temperature, which degraded the HZO film~\cite{Asapu2022FrontMat}. Additionally, the co-doped stack with the SiON/HZO interface displays slightly higher leakage in the positive direction and breaks down earlier than the other two co-doped stacks. This observation aligns with the results seen in FM, indicating that the increased leakage current in both stacks contributes to their reduced endurance compared to others.\\
Considering that the 2P$_r$ values from FM could be significantly influenced by leakage current, the P-V loops of each stack were extracted from PUND measurements to compare their actual P$_r$ values, as shown in Fig.~\ref{PE_loops}a. Although the stack with mono-doped HZO initially exhibits the largest P$_r$, it quickly diminishes during cycling, closing by 10$^3$ cycles. Consistent with the FM results, the mono-doped HAO displays a much smaller P$_r$ but shows almost no degradation up to 10$^4$ cycles. The co-doped stack with the SiON/HZO interface combines the disadvantages of the both former stacks: rapid degradation and low P$_r$. In contrast, both co-doped stacks with the SiON/HAO interface demonstrate outstanding P-V characteristics, merging the advantages of mono-doped HZO and HAO: high P$_r$ and no degradation up to 10$^4$ cycles. Furthermore, the J–V characteristics, excluding the contributions from displacement and leakage currents, were extracted from the PUND measurements, as shown in Fig.~\ref{PE_loops}b. Compared to other stacks, the SiON/HAO/HAO/HZO stack exhibits a significantly more concentrated switching current, which suggests the best ferroelectric switching behavior. Notably, after cycling, a negative current peak is observed under positive bias in both stacks with an SiON/HZO interface—this is an artifact resulting from the leaky ferroelectric films during the PUND measurements, as shown in the inset of Fig.~\ref{PE_loops}b. Ultra-high leakage without switching current can be noticed under positive bias.\\
The smaller coercive voltages observed in both the Al mono-doped and SiON/HAO/HAO/HZO co-doped films are likely related to their higher o\{200\} phase fractions, as shown in FIG.~\ref{XRD_Comparison}. Additionally, the imprint in the capacitor with the HZO/TiN stack is slightly reduced, likely due to the introduction of oxygen vacancies by Al doping at the bottom of the ferroelectric layer, which helps counteract the built-in field arising from the asymmetric MFIS stack.\\
Therefore, it can be concluded that Zr/Al heterogeneous co-doping significantly enhances the endurance and 2P$_r$ of MFIS stacks, compared to mono-doped HZO and HAO films.\\
However, the SiON/HZO interface should be avoided in both mono- and co-doped HfO$_2$ films, as it results in a significantly degraded performance. This degradation is likely due to the diffusion of Zr\textsuperscript{4+} into the SiON insulator layer, as indicated by the ToF-SIMS profiles in Fig.~\ref{Diffusion}. The presence of diffused Zr\textsuperscript{4+} ions in the SiON layer likely deteriorates the quality of the IL, leading to higher leakage and consequently faster degradation of the entire stack.\\
In conclusion, the deliberate design of Zr/Al dopant profiles in HfO$_2$ MFIS stacks offers a powerful route for modulating crystallization pathways and phase distributions. By fine-tuning the spatial arrangement of these dopants, we achieve enhanced endurance and remanent polarization in FeFET gate stacks, demonstrating the practical benefits of heterogeneous co-doping beyond single-dopant strategies. Notably, avoiding direct Zr/SiON interfaces is critical for suppressing leakage and maximizing device reliability.\\
\begin{acknowledgments}
The authors gratefully acknowledge the valuable support of their colleagues: Lisa Roy for her assistance with the GIXRD measurements, and Kati Biedermann and Jennifer Salah Emara for their contributions to the ToF-SIMS measurements. This work has received funding from German Bundesministerium f\"{u}r Bildung und Forschung (BMBF) through
the project T4T under grant agreement No. 16ME0483.\\
\end{acknowledgments}
\section*{Conflict of Interest}
The authors have no conflicts to disclose.
\section*{data availability}
The data that support the findings of this study are available from the corresponding author upon reasonable request.
\nocite{*}
\section*{References}
\bibliography{References.bib}

\end{document}